\documentclass[aps,prd,groupedaddress,showpacs]{revtex4}

\usepackage{epsfig}
\usepackage{graphicx}
\usepackage{amsmath}
\usepackage{amsfonts,amssymb}
\usepackage{color}

\newtheorem{theorem}{Theorem}

\def\endprf{\hfill  {\vrule height6pt width6pt depth0pt}\medskip}

\begin{document}


\title{Exact solution of Dyson-Schwinger equations for a scalar field theory}


\author{Marco Frasca}
\email[e-mail:]{marcofrasca@mclink.it}
\affiliation{Via Erasmo Gattamelata, 3 \\
             00176 Roma (Italy)}


\date{\today}

\begin{abstract}
We exactly solve Dyson-Schwinger equations for a massless quartic scalar field theory. n-point functions are computed till n=4 and the exact propagator computed from the two-point function. The spectrum is so obtained, being the same of a harmonic oscillator. Callan-Symanzik equation for the two-point function gives the beta function. This gives the result that this theory has only trivial fixed points. In the low-energy limit the coupling goes to zero making the theory trivial and, at high energies, it reaches infinity. No Landau pole appears, rather this should be seen as a precursor, in a weak perturbation expansion, of the coupling reaching the trivial fixed point at infinity. Using a mapping theorem, recently proved, between massless quartic scalar field theory and gauge theories, we derive some properties of the latter.
\end{abstract}


\maketitle


\section{Introduction}

Dyson-Schwinger equations are a pivotal technique to manage a quantum field theory when perturbation methods fail. Indeed, there exists a vast literature of their application in the area of quantum chromodynamics producing a lot of activity and motivating people to significant efforts in lattice computations \cite{as,cr}. Indeed, these studies are underway yet and a clear and commonly accepted picture is not emerged so far. The main difficulty in managing this kind of equations relies on the fact that at a given order a Dyson-Schwinger equation is overdetermined depending on the solution of higher order equations. This makes such equations in need for a truncation scheme that can make them manageable. Anyhow, they represent a beautiful example of exact results in quantum field theory.

On this basis, it is of great importance to obtain exact solutions for them and this is our aim. Such solutions are very difficult to find for the reasons given above and it is surely of some relevance to get someone, to have an understanding of what is going on in this case. The theory we consider here is that of a massless quartic scalar field. This theory is so important to be a relevant example for whatever technique one needs to exploit in quantum field theory. The reason is that a wealth of results are known about \cite{pes}.

Indeed, quite recently, we proved a theorem that a map exists between such a field and a Yang-Mills theory at a classical level \cite{fra3}. The proof of the theorem has been criticized by Terence Tao, showing that it was in need of a deep revision. This was accomplished quite recently \cite{fra4} producing the conclusion that these theories map at a perturbative level when the coupling is taken to be infinitely large \cite{fra5}. The relevance of this theorem relies on the possibility to move the results obtained for a theory, easier to manage, to the other in the given limits.

To exploit this opportunity, one needs how to solve classical equations of motion for the scalar field theory. We did this quite recently \cite{fra0} but another approach was devised for this aim in our earlier publications \cite{fra6,fra7}. In this way, a quantum field theory can be built with them and these solutions map on the Yang-Mills theories for a very large coupling. A peculiar property that these solutions have is that they describe massive wave-like motions of the field notwithstanding the field itself has no mass from the start. Indeed, mass emerges from field self-interaction. We will recover all the conclusions drawn in \cite{fra0} starting from Dyson-Schwinger equations.

The paper is so structured. In sec. \ref{ds} we give the Dyson-Schwinger equations for a massless quartic scalar field. In sec. \ref{sds} we present their solution. In sec. \ref{rct} we analyze the question of the running coupling and the triviality appearing at low energies for this theory. In sec. \ref{map} we exploit the mapping between scalar field theory and Yang-Mills theories drawing some conclusions about the latter. Finally, in sec. \ref{cn} conclusions are presented.

\section{Dyson-Schwinger equations for a scalar field theory\label{ds}}

We consider the simplest interacting field theory with action
\begin{equation}
    S=\int d^4x\left[\frac{1}{2}(\partial\phi)^2-\frac{\lambda}{4}\phi^4\right]
\end{equation}
and we can immediately write down the generating functional for the quantum field theory as
\begin{equation}
    Z[j]=N\int[d\phi]\exp\left[iS[\phi]+i\int d^4xj\phi\right]
\end{equation}
having chosen $N$ so that $Z[0]=1$. Corresponding Dyson-Schwinger equations can be obtained through techniques showed in \cite{alk,ben}. These are given by, after currents are set to zero,
\begin{eqnarray}
    &&-\partial^2G_1(x)+\lambda G_1^3(x)+3\lambda G_1(x)G_2(0)+\lambda G_3(0,0)=0 \\ \nonumber
    &&\\ \nonumber
    &&-\partial^2G_2(x-y)+3\lambda\left[G_1^2(x)+G_2(0)\right]G_2(x-y) \\ \nonumber
    &&+3\lambda G_1(x)G_3(0,x-y)+\lambda G_4(0,0,x-y)=i\delta^4(x-y) \\ \nonumber
    &&\\ \nonumber
    &&-\partial^2G_3(x-y,x-z)+\lambda\left[6G_1(x)G_2(x-y)G_2(x-z)+3G_1^2(x)G_3(x-y,x-z)\right. \\ \nonumber
    &&+3G_2(x-z)G_3(0,x-y)+3G_2(x-y)G_3(0,x-z) \\ \nonumber
    &&\left.+3G_2(0)G_3(x-y,x-z)+3G_1(x)G_4(0,x-y,x-z)+G_5(0,0,x-y,x-z)\right]=0 \\ \nonumber
    &&\\ \nonumber
    &&-\partial^2G_4(x-y,x-z,x-w)+\lambda\left[6G_2(x-y)G_2(x-z)G_2(x-w)
    \right. \\ \nonumber
    &&+6G_1(x)G_2(x-y)G_3(x-z,x-w)+6G_1(x)G_2(x-z)G_3(x-y,x-w)\\ \nonumber
    &&+6G_1(x)G_2(x-w)G_3(x-y,x-z)+3G_1^2(x)G_4(x-y,x-z,x-w) \\ \nonumber
    &&+3G_2(x-y)G_4(0,x-z,x-w)+3G_2(x-z)G_4(0,x-y,x-w)  \\ \nonumber
    &&+3G_2(x-w)G_4(0,x-y,x-z)+3G_2(0)G_4(x-y,x-z,x-w) \\ \nonumber
    &&\left.+3G_1(x)G_5(0,x-y,x-z,x-w)+G_6(0,0,x-y,x-z,x-w)\right]=0 \\ \nonumber
    &\vdots&
\end{eqnarray}
being
\begin{equation}
   G_n(x_1,x_2,\ldots,x_n)=(-i)^n\frac{\delta^n \ln(Z)}{\delta j(x_1)\delta j(x_2)\ldots\delta j(x_n)}
\end{equation}
n-point functions. Our aim in the next section will be to compute all of them through 1-point and 2-point functions solving completely all the set of equations.

\section{Exact solution of Dyson-Schwinger equations\label{sds}}

The above set of equations can be solved straightforwardly. Let us consider the first Dyson-Schwinger equation
\begin{equation}
    -\partial^2G_1(x)+\lambda G_1^3(x)+3\lambda G_1(x)G_2(0)+\lambda G_3(0,0)=0.
\end{equation}
Imposing the conditions $G_2(0)=0$ and $G_3(0,0)=0$, this reduces to
\begin{equation}
    -\partial^2G_1(x)+\lambda G_1^3(x)=0
\end{equation}
that is $G_1(x)$ is an exact solution of the equation of motion of the scalar field. We emphasize here that this is an ad hoc choice for a particular class of exact solutions of this equation. We know a set of these solutions given by \cite{fra0}:
\begin{equation}
    G_1(x)=\mu\left(\frac{2}{\lambda}\right)^{\frac{1}{4}}{\rm sn}(p\cdot x+\theta,i)
\end{equation}
being sn a Jacobi elliptical function, and $\mu$ and $\theta$ two integration constants. This solution holds with the dispersion relation
\begin{equation}
\label{eq:disp}
    p^2=\mu^2\sqrt{\frac{\lambda}{2}}
\end{equation}
showing that these solutions represent massive excitations of the field, notwithstanding we started with a massless field. We will be able to fix the value of $\theta$ and derive the spectrum of the theory in agreement with ref.\cite{fra0}. We note that this solution retains translational invariance as it can always be rewritten as
\begin{equation}
    G_1(x-y)=\mu\left(\frac{2}{\lambda}\right)^{\frac{1}{4}}{\rm sn}(p\cdot (x-y)+\theta,i)
\end{equation}
and being a solution yet of the equation we started from. So, for the following we will use this solution written down in a translational-invariant form.

Then, the second equation can be written as
\begin{eqnarray}
\label{eq:2pf}
&&-\partial^2G_2(x-y)+3\lambda\left[G_1^2(x-y)+G_2(0)\right]G_2(x-y) \\ \nonumber
    &&+3\lambda G_1(x-y)G_3(0,x-y)+\lambda G_4(0,0,x-y)=i\delta^4(x-y)
\end{eqnarray}
and setting
\begin{eqnarray}
     G_3(0,x-y)&=& 0 \\ \nonumber
     G_4(0,0,x-y) &=& 0
\end{eqnarray}
we arrive at
\begin{equation}
-\partial^2G_2(x-y)+3\lambda G_1^2(x-y)G_2(x-y)=i\delta^4(x-y).
\end{equation}
In order to solve this equation we consider the corresponding one
in the rest frame with ${\bf p}=0$ and $p_0=\mu(\lambda/2)^{\frac{1}{4}}$, that is we put the gradients to zero. One gets
\begin{equation}
\label{eq:g2}
   G_2(x-y)=-i\delta^3(x-y)\theta(t_x-t_y)\frac{1}{\mu(2^3\lambda)^{\frac{1}{4}}}
   {\rm cn}\left[\left(\frac{\lambda}{2}\right)^{\frac{1}{4}}\mu (t_x-t_y)+\theta_n,i\right]
   {\rm dn}\left[\left(\frac{\lambda}{2}\right)^{\frac{1}{4}}\mu (t_x-t_y)+\theta_n,i\right]
\end{equation}
when the phase is taken to be $\theta_k$, being ${\rm cn}(\theta_k,i)=0$
fixing in this way the phase in $G_1(x)$. These phases grant the above to be an exact solution of eq.(\ref{eq:2pf}) and that $G_2({\bf x}-{\bf y},0)=0$ so, taking the limit properly, one can assume $G_2(0)=0$. This result can also be obtained through a proper regularization scheme for the theory.

Now, we are able to get the propagator in momentum space. So, let us rewrite the above formula as \cite{gra}
\begin{eqnarray}
   G_2(x-y)&=&-i\delta^3(x-y)\theta(t_x-t_y)\frac{1}{\mu^2(2\lambda)^{\frac{1}{2}}}
   \frac{d}{dt} {\rm sn}\left[\left(\frac{\lambda}{2}\right)^{\frac{1}{4}}\mu (t_x-t_y)+\theta_n,i\right] \nonumber \\
   &=&-i\delta^3(x-y)\theta(t_x-t_y)\frac{1}{(8\lambda)^{\frac{1}{4}}\mu}
   \frac{\pi^2}{K(i)^2}\sum_{n=0}^\infty(2n+1)(-1)^n
   \frac{e^{-\left(n+\frac{1}{2}\right)\pi}}{1+e^{-(2n+1)\pi}}\times \\ \nonumber
   &&\cos\left[(2n+1)\frac{\pi}{2K(i)}\left(\frac{\lambda}{2}\right)^\frac{1}{4}\mu (t_x-t_y)
   +\chi_{n,k}\right].
\end{eqnarray} 
being $\chi_{n,k}=(2n+1)\frac{\pi}{2K(i)}\theta_k$. The simplest choice could be $\theta_{k=0}=K(i)$ being $K(i)=\int_0^{\frac{\pi}{2}}d\theta/\sqrt{1+\sin^2\theta}$ an elliptic integral (but these phases are generally complex numbers). So, we get
\begin{eqnarray}
   G_2(x-y)&=&i\delta^3(x-y)\theta(t_x-t_y)\frac{1}{(8\lambda)^{\frac{1}{4}}\mu}\frac{\pi^2}{K(i)^2}\sum_{n=0}^\infty(2n+1)
   \frac{e^{-\left(n+\frac{1}{2}\right)\pi}}{1+e^{-(2n+1)\pi}}\times \\ \nonumber
   &&\sin\left[(2n+1)\frac{\pi}{2K(i)}\left(\frac{\lambda}{2}\right)^\frac{1}{4}\mu (t_x-t_y)\right].
\end{eqnarray} 
Already at this stage, we are able to identify the theory spectrum as given by
\begin{equation}
\label{eq:spe1}
   m_n=(2n+1)\frac{\pi}{2K(i)}\left(\frac{\lambda}{2}\right)^\frac{1}{4}\mu
\end{equation}
being this the one of a harmonic oscillator. This is the spectrum of the quantum field theory and we note as the bare mass given in eq.(\ref{eq:disp}) is renormalized. This is seen setting $n=0$ in eq.(\ref{eq:spe1}) and noting a finite renormalization constant given by $\frac{\pi}{2K(i)}$.

We can draw the conclusion from this analysis that this gives the exact propagator of the theory. Defining the propagator through the two-point function as
\begin{equation}
   \Delta(x-y)=\theta(t_x-t_y)G_2(x-y)+\theta(t_y-t_x)G_2(y-x),
\end{equation}
we can write it down, after a Lorentz boost, as
\begin{equation}
\label{eq:prop}
    \Delta(p)=\sum_{n=0}^\infty\frac{B_n}{p^2-m_n^2+i\epsilon}
\end{equation}
being
\begin{equation}
    B_n=(2n+1)^2\frac{\pi^3}{4K^3(i)}\frac{e^{-(n+\frac{1}{2})\pi}}{1+e^{-(2n+1)\pi}}.
\end{equation}
We note that the propagator is in agreement with K\"allen-Lehman form and has all the coefficients being greater than zero. Further, it reduces in the proper limit of the propagator of the free theory when $\lambda$ is taken to be zero being $\sum_{n=0}^\infty B_n=1$. In an infrared expansion in the inverse of the coupling this represents the next to leading order correction \cite{fra0}.

The equation for the three-point function is
\begin{equation}
-\partial^2G_3(x-y,x-z)+3\lambda G_1^2(x-y)G_3(x-y,x-z)=-6\lambda G_1(x-y)G_2(x-y)G_2(x-z) 
\end{equation}
where we have set
\begin{eqnarray}
   G_4(0,x-y,x-z)&=&0 \\ \nonumber 
   G_5(0,0,x-y,x-z) &=& 0 
\end{eqnarray}
to be checked {\sl a posteriori}. The solution is straightforwardly obtained as
\begin{equation}
   G_3(x-y,x-z)=-6\lambda\int dx_1 G_2(x-x_1)G_1(x_1-y)G_2(x_1-y)G_2(x_1-z) 
\end{equation}
and it is easy to verify that $G_3(0,x-z)=G_3(x-y,0)=0$ using the property of Heaviside function $\theta(x)\theta(-x)=0$.

Finally, we consider the four-point function. The corresponding Dyson-Schwinger equation is
\begin{eqnarray}
&&-\partial^2G_4(x-y,x-z,x-w)+3\lambda G_1^2(x-y)G_4(x-y,x-z,x-w)= \\ \nonumber
&&-6\lambda G_2(x-y)G_2(x-z)G_2(x-w)-6\lambda G_1(x-y)G_2(x-y)G_3(x-z,x-w) \\ \nonumber
&&-6\lambda G_1(x-y)G_2(x-z)G_3(x-y,x-w)-6\lambda G_1(x-y)G_2(x-w)G_3(x-y,x-z)
\end{eqnarray}
where we have set
\begin{eqnarray}
    G_5(0,x-y,x-z,x-w)&=& 0 \\ \nonumber
    G_6(0,0,x-y,x-z,x-w)&=& 0.
\end{eqnarray}
The solution is easily written down as
\begin{eqnarray}
    &&G_4(x-y,x-z,x-w)=-6\lambda\int dx_1 G_2(x-x_1)G_2(x_1-y)G_2(x_1-z)G_2(x_1-w) \\ \nonumber
    &&-6\lambda\int dx_1G_2(x-x_1)\left[G_1(x_1-y)G_2(x_1-y)G_3(x_1-z,x_1-w)\right. \\ \nonumber
    &&\left.+G_1(x_1)G_2(x_1-z)G_3(x_1-y,x_1-w)
    +G_1(x_1-y)G_2(x_1-w)G_3(x_1-y,x_1-z)\right].
\end{eqnarray}
Again, it is easy to check {\sl a posteriori} that $G_4(0,x-y,x-z)=0$ showing that our solution to Dyson-Schwinger equations is indeed correct.

So, one can iterate this procedure to any order and obtain the full set of n-point functions. We see from this that the theory is exactly solved having all the n-point functions and the spectrum.

The choice of the exact solution $G_1$ is arbitrary. This means that one can build a full theory starting e.g. from instantons. This makes this approach quite general. But, whatever is the value of $G_1$, one must be able to solve the equation for the two-point function and this is not generally known.

\section{Running coupling and triviality\label{rct}}

Having the exact propagator of the theory, our aim is to compute the beta function. We want to see if this theory is trivial as already happens for dimensions greater than four \cite{aiz}. It is already known that this theory, in the limit of a weak coupling, has an energy scale for which small perturbation theory fails and this fact is normally identified through a Landau pole in the beta function \cite{pes}. Our aim can be easily reached by computing Callan-Symanzik equation from the propagator (\ref{eq:g2}). This yields immediately
\begin{equation}
   \mu\frac{\partial G_2}{\partial\mu}-4\lambda\frac{\partial G_2}{\partial\lambda}
   +2\gamma G_2=0
\end{equation}
being $\gamma=0$. It is easy to read from this equation that
\begin{equation}
   \beta(\lambda)=-4\lambda
\end{equation}
that gives the following running coupling
\begin{equation}
   \lambda(p)=\lambda_0\frac{p^4}{\Lambda^4}.
\end{equation}
This is in agreement with recent analysis \cite{sus1,sus2}.
From this we learn that this theory has only trivial fixed points: the coupling goes to zero in the infrared and goes to infinity in the ultraviolet. Landau pole is not seen, rather it represents a failure of small perturbation theory. This just gives a clue of the fact that a theory has a trivial fixed point to infinity. The main conclusion here is that this theory becomes trivial in the infrared as the coupling reaches zero making the theory free. In the same limit, the theory develops a mass gap as can be seen from the spectrum of the theory.  

\section{Mapping theorem and Yang-Mills theories\label{map}}

The relevance of this scalar theory in the understanding of gauge theories has emerged with a proof of a mapping theorem given in \cite{fra3}. The proof was criticized by Terence Tao as he pointed out that a selected set of solutions of the classical equation of the scalar field is not granted to be an extremum on all the space of connections for the classical gauge theory. This criticism was successfully answered in \cite{fra4} and the theorem is proved in the form we will give below \cite{fra5}. It should be emphasized the relevance of Tao's criticism to give a conclusive understanding of the mapping between Yang-Mills fields and scalar field theories.

So, we can state \cite{fra3,fra4,fra5}:
\begin{theorem}[Mapping]
\label{teo1}
An extremum of the action
\begin{equation}
    S = \int d^4x\left[\frac{1}{2}(\partial\phi)^2-\frac{\lambda}{4}\phi^4\right]
\end{equation}
is also an extremum of the SU(N) Yang-Mills Lagrangian when one properly chooses $A_\mu^a$ with some components being zero and all others being equal, and $\lambda=Ng^2$, being $g$ the coupling constant of the Yang-Mills field, when only time dependence is retained. In the most general case the following mapping holds
\begin{equation}
    A_\mu^a(x)=\eta_\mu^a\phi(x)+O(1/g)
\end{equation}
being $\eta_\mu^a$ constant, that becomes exact for the Lorenz gauge.
\end{theorem}

The relevance of this theorem relies on the fact that, in the limit $g\rightarrow\infty$ for the Yang-Mills fields, all the results obtained so far for the scalar field apply immediately to the gauge fields. From this one gets the propagator, n-point functions and the spectrum. But, it is really striking to see how the behavior of the running coupling fails in the ultraviolet where the above theorem cannot be applied and here one has asymptotic freedom. So, one can conclude that the running coupling for a Yang-Mills field attains a maximum at intermediate energies, as it goes to zero at both ends of the range and this is what is seen in lattice computations \cite{lat1,lat2,lat3,lat4}.

Finally, we write here the two-point function, in the limit $g\rightarrow\infty$ and for the Landau gauge, given by
\begin{equation}
   \tilde D_{\mu\nu}^{ab}(p^2)=\delta_{ab}\left(g_{\mu\nu}
   -\frac{p_\mu p_\nu}{p^2}\right)\sum_{n=0}^\infty\frac{iB_n}{p^2-m_n^2+i\epsilon}
\end{equation}
but this represent a correction to the propagator of the theory in the infrared at order $O(1/Ng^2)$ in agreement with our discussion above while a correction to order $O(1/\sqrt{Ng^2})$ is given in \cite{fra8}. Here is
\begin{equation}
    B_n=(2n+1)^2\frac{\pi^3}{4K^3(i)}\frac{e^{-(n+\frac{1}{2})\pi}}{1+e^{-(2n+1)\pi}} 
\end{equation}
where use has been made of eq.(\ref{eq:prop}). Here, for SU(N),
\begin{equation}
   m_n=(2n+1)\frac{\pi}{2K(i)}\left(\frac{Ng^2}{2}\right)^\frac{1}{4}\mu.
\end{equation}
It is easy to check that
\begin{equation}
   \tilde D^{ab}_{\mu\nu}(p^2)\rightarrow\frac{i}{p^2}
\end{equation}
when the limit of momentum going to infinity is taken. Similarly, for the running coupling one has
\begin{equation}
    \alpha_s(p)=\alpha_s^0\frac{p^4}{\Lambda^4}
\end{equation}
to hold in the same limits as above. These appear to agree with lattice computations \cite{lat1,lat2,lat3,lat4}. 

The main conclusion of this section is that a Yang-Mills theory, with SU(N) gauge group, has only trivial fixed points as the running coupling reaches zero at both ends of the energy range.
   
\section{Conclusions\label{cn}}

We have shown how, for a quartic massless scalar field, Dyson-Schwinger equations can be exactly solved. We have seen the way such a field gets a mass gap and has trivial fixed points. Particularly, at lower momenta, coupling goes to zero making the theory trivial. A Landau pole is not observed, rather this should represent a failure of the perturbation method and a clue that a theory has a trivial fixed point at infinity.

It is interesting to note that, all the theories present in the Standard Model have only trivial fixed points. This situation may change for supersymmetry and string theory as well.

Relevant conclusions are obtained for Yang-Mills theories from this exact solution. These are given by a mapping theorem recently proved. Our hope is that our solutions, currently in agreement with results emerging from lattice computations, could help to clarify the situation at hand.

\newpage



\end{document}